\documentclass[usegraphicx]{mn2e}

\title{Light Curve Morphology Study of UW CrB -- Evidence for a 5 d Superorbital Period}
\author[Hakala, Hjalmarsdotter, Hannikainen \& Muhli]{Pasi Hakala$^{1}$\thanks{E-mail:
pahakala@utu.fi},
Linnea Hjalmarsdotter,$^{2}$
Diana C. Hannikainen$^{2,3}$
Panu Muhli$^{2}$\thanks{Based on observations taken with the Nordic Optical Telescope,
La Palma.}\\
$^{1}$Tuorla Observatory, V\"ais\"al\"antie 20, University Of Turku, FIN-21500 Piikki\"o, Finland. \\
$^{2}$Observatory, P.O. BOX 14, University of Helsinki, FIN-00014 University of Helsinki, Finland.\\
$^{3}$Mets\"ahovi Radio Observatory/TKK, Mets\"ahovintie 114, FIN-02540 Kylm\"al\"a, Finland.}
\begin{document}

\date{}

\pagerange{\pageref{firstpage}--\pageref{lastpage}} \pubyear{2002}

\maketitle

\label{firstpage}

\begin{abstract}
Since its discovery in 1990, UW CrB (also known as MS1603+2600) has
remained a peculiar source without firm classification. Our current understanding is that it is an Accretion
Disc Corona (ADC) low mass X-ray binary. In this paper we present results from our 
photometric campaign dedicated to studying the changing morphology of the optical light curves. We
find that the optical light curves show remarkable evidence for strongly evolving light curve shapes. In addition we find that these changes show a modulation at a period of $\sim$ 5 days. We interpret these changes as either due to strong periodic accretion disc warping or other geometrical changes due to disc precession at a period of 5 days. Finally, we have detected 11 new optical bursts, the phase distribution of which supports the idea of a vertically extended asymmetric accretion disc.
\end{abstract}


\section{Introduction}

UW CrB (also known as MS1603+2600) was discovered in X-rays during the \textit{Einstein} medium
sensitivity survey (Morris et al. 1990). Morris et al. (1990) also identified the optical
counterpart of the X-ray source. A star with blue, accretion disc-like spectrum with Balmer emission lines was found close to the X-ray position. Subsequent optical photometry revealed a period of 111 minutes. Morris et al. (1990) also noted that  the shape of the optical light curve seemed to change from one night to the other. The system is located at a high galactic latitude of 47$^o$. This, together with its low X-ray flux of 1.14$\times$10$^{-12}$ erg/s (Morris et al. 1990), suggests that either the source is underluminous in X-rays for a low mass X-ray binary (LMXB) or that it is actually located in the galactic halo. 

Earlier models for UW CrB  have involved both LMXB and magnetic CV scenarios (Morris et al. 1990, Ergma \& Vilhu 1993). Based on evolutionary calculations Ergma \& Vilhu prefer the LMXB option. Hakala et al. (1998) 
compared the ROSAT spectra of UW CrB with soft X-ray spectra of various classes of interacting binaries. Their conclusion 
was that the only class where the spectral fits seemed to match those of the source was that of the soft X-ray transients (SXT) in quiescence. This could also explain the relatively low \textit{F$_{X}$/F$_{opt}$} ratio the system has. The main complication for this explanation arises from the fact that, according to our knowledge, the optical light curves of other SXT's do not vary like they do in UW CrB. The same is true for the magnetic CV hypothesis (AM Herculis type). In addition, Hakala et al. (1998)
report a negative result from their search for circular polarisation.

Hakala et al. (1998) also published the results of further optical photometry together with the ROSAT observations. They confirmed the extreme variability seen in the optical light curve pulse shapes. Even if there are dramatic changes  in the optical light curve shape, the overall optical flux level does NOT seem to vary much. The analysis of ROSAT data proved that the soft X-ray spectrum of UW CrB can be modelled with a single blackbody component with a temperature of  0.24 keV. The ROSAT data also showed that there is no significant interstellar absorption in the direction of the source. 

Mukai et al. (2001) presented ASCA observations of UW CrB. They detect a single type I X-ray burst, which rules out a black hole as the primary component in the system (which was suggested by Hakala et al. 1998).
These bursts have also been seen in the optical (Muhli et al. 2004, Hynes et al. 2004.). 
Mukai et al. (2001) propose that UW CrB is a short period X-ray dipper, like 4U1916-05. They also note that the system would probably in this case have to reside in the outer galactic halo, and could have formed in the globular cluster Palomar 14, which is located 11$^o$ from UW CrB at a distance of 73.8 kpc. The {\it Chandra} observations of UW CrB by
Jonker et al.(2003) modelled the X-ray spectrum with a simple powerlaw of spectral index $\alpha$= 2.0. They conclude that the source is an accretion disc corona (ADC) system at a distance of 11--24 kpc.
Finally, two {\it XMM-Newton} observations, separated by 2.5 days, have shown that in addition to the
changes in the shape of the optical light curve, also the X-ray and UV light curves change significantly on a time scale of days (Hakala et al. 2005). They also detect several X-ray bursts confirming the neutron star nature of the primary. The overall {\it XMM-Newton} EPIC X-ray spectrum cannot be fitted with a simple model. In addition
to the blackbody component and the powerlaw used to model the {\it ROSAT} and {\it Chandra} spectra Hakala
et al. (2005) find that an additional thermal plasma (MEKAL) component is required in order to fit the spectrum.  The phase resolved X-ray spectra suggest that the most likely cause for the X-ray modulation is changing partial covering of the X-ray emission components over the orbital cycle. This strongly suggests that 
UW CrB is an ADC source with a thick or warped asymmetric accretion disc.

In this paper we will present further (high time resolution) photometry of UW CrB. We will discuss the observations, followed by a statistical analysis of our data. We will also analyse the 11 optical bursts detected in our data set, together with the bursts detected earlier in various data sets (Hynes et al. 2004, Hakala et al. 2005). Finally we discuss the possible origin of the modulation detected and its implication on our understanding of UW CrB, and accretion discs in LMXB's in general.   

\section{Observations}

We observed UW CrB on the Nordic Optical Telescope (NOT), La Palma, on four different occasions.  The system setup was identical during these runs. NOT was used with a Cassegrain spectrograph/imager ALFOSC, which in turn was equipped with a Loral-Lesser, thinned, AR coated, back illuminated 2048 by 2048 CCD detector. This detector has a relatively flat quantum efficiency of more than 80\% all the way from B to I band. In order to increase the time resolution, the chip was subwindowed and binned by a factor of two resulting in a pixel size of 0.36". As a result we were able to obtain time series photometry at a true time resolution of 13 s (10 s exposure time and 3 s readout overhead). All the observations were done in white light in order to maximize the S/N. All the data were reduced in a normal manner for CCD images (i.e. bias subtraction and flat field correction). The field typically included at least two other stars. On all occasions we used the ``bright", about 14th mag star, located $\sim$ 30" NNE of  of UW CrB (star ``C", see Hakala et al. 1998 for a finding chart) as a comparison star. 

\begin{table}
\caption{Observing log. All data were obtained in white light with a typical time resolution
of 10-15 s.}
\begin{center}
\begin{tabular}{|c|c|c|}
\hline
Start time (HJD) & Data set & Duration (d) \\   
\hline
2451212.705 & Feb 1999 & 0.083 \\
2451213.727 & Feb 1999 & 0.077 \\
2451214.669 & Feb 1999 & 0.125 \\
2451215.676 & Feb 1999 & 0.124 \\
\hline
2451304.561 & May 1999 & 0.178 \\
2451305.562 & May 1999 & 0.176 \\
2451306.572 & May 1999 & 0.162 \\
2451307.569 & May 1999 & 0.162 \\
2451308.559 & May 1999 & 0.162 \\
2451309.577 & May 1999 & 0.154 \\
\hline
2452455.386 & Jul 2002 & 0.172 \\
2452456.378 & Jul 2002 & 0.176 \\
2452457.452 & Jul 2002 & 0.105 \\
2452458.377 & Jul 2002 & 0.157 \\
2452459.378 & Jul 2002 & 0.077 \\
2452460.376 & Jul 2002 & 0.175 \\
2452461.379 & Jul 2002 & 0.171 \\
2452462.383 & Jul 2002 & 0.169 \\
\hline
2452492.379 & Aug 2002 & 0.086 \\
\hline
\end{tabular}
\end{center}
\label{default}
\end{table}%

Our first data set was obtained in February 1999, and covers data from four consecutive 
nights. The second data set, obtained  in May 1999, consists of six consecutive half nights of data. The source was
observed for about two orbital cycles each half night, enabling the construction of a phase folded light curve each night.
The light curves from May 1999 (folded into 100 phase bins) are plotted in Figure 1. Each subplot contains data from 
a single night. Our third data set (from June/July 2002) consists of eight consecutive half-nights of observations. As mentioned above, the instrument setup was identical and also the same comparison star was used. These data are plotted in Figure 2.
We typically have about 1000 photometric points per night, which means that each of the phase bins in the folded plots (Figures 1 \& 2) is an average of ten measurements (i.e. about 100 s effective exposure time).

\begin{figure*}
% \centerline{\psfig{figure=lc1999.ps,height=20.0,width=18.0,angle=0}}
\includegraphics[scale=0.7]{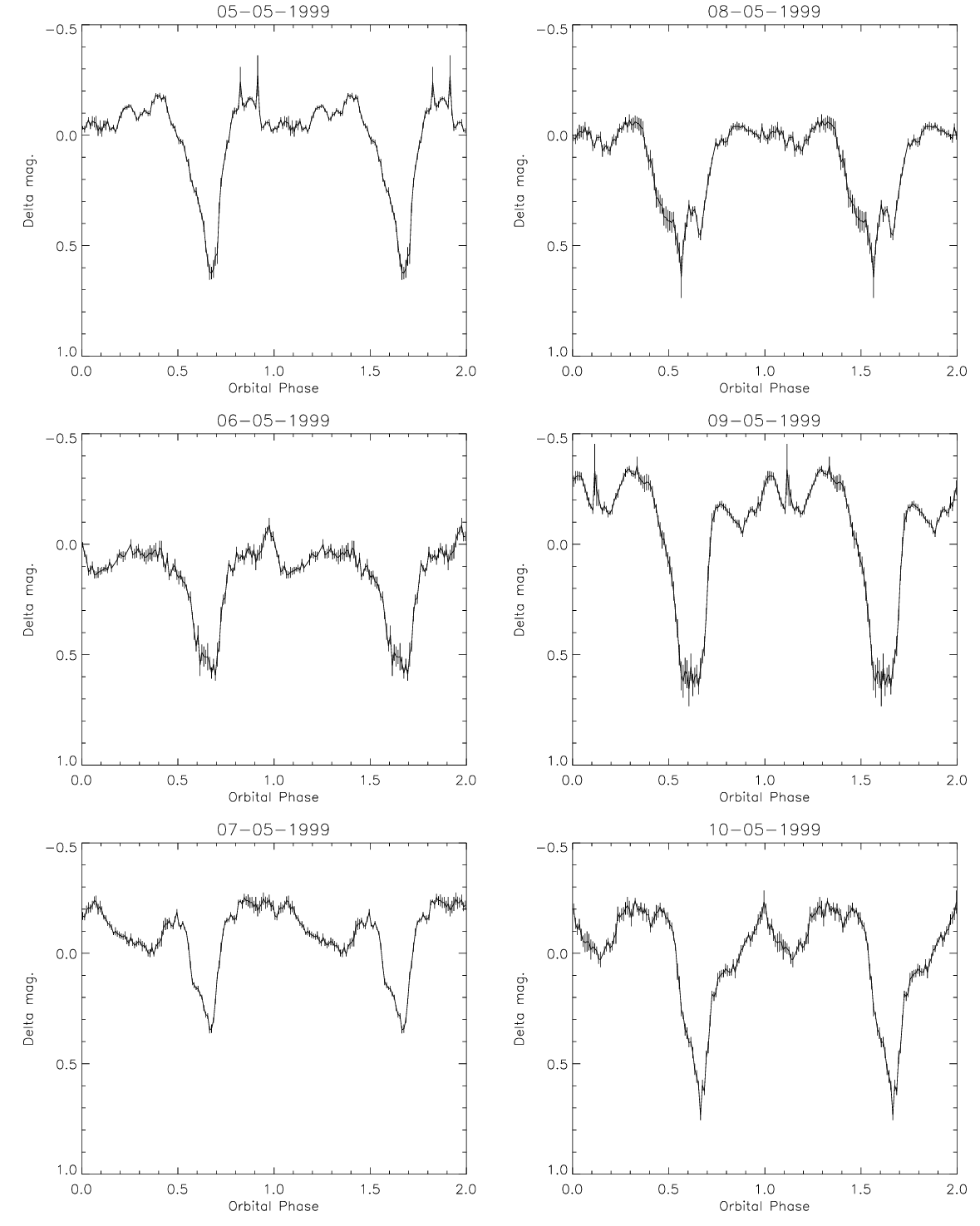}
 \caption{The May 1999 data set. Subpanels show data from different nights (as indicated) folded into 100 phase bins and plotted twice for clarity. The orbital phase is arbitrary}
\end{figure*}

\begin{figure*}
% \centerline{\psfig{figure=lc1999.ps,height=20.0,width=18.0,angle=0}}
\includegraphics[scale=0.7]{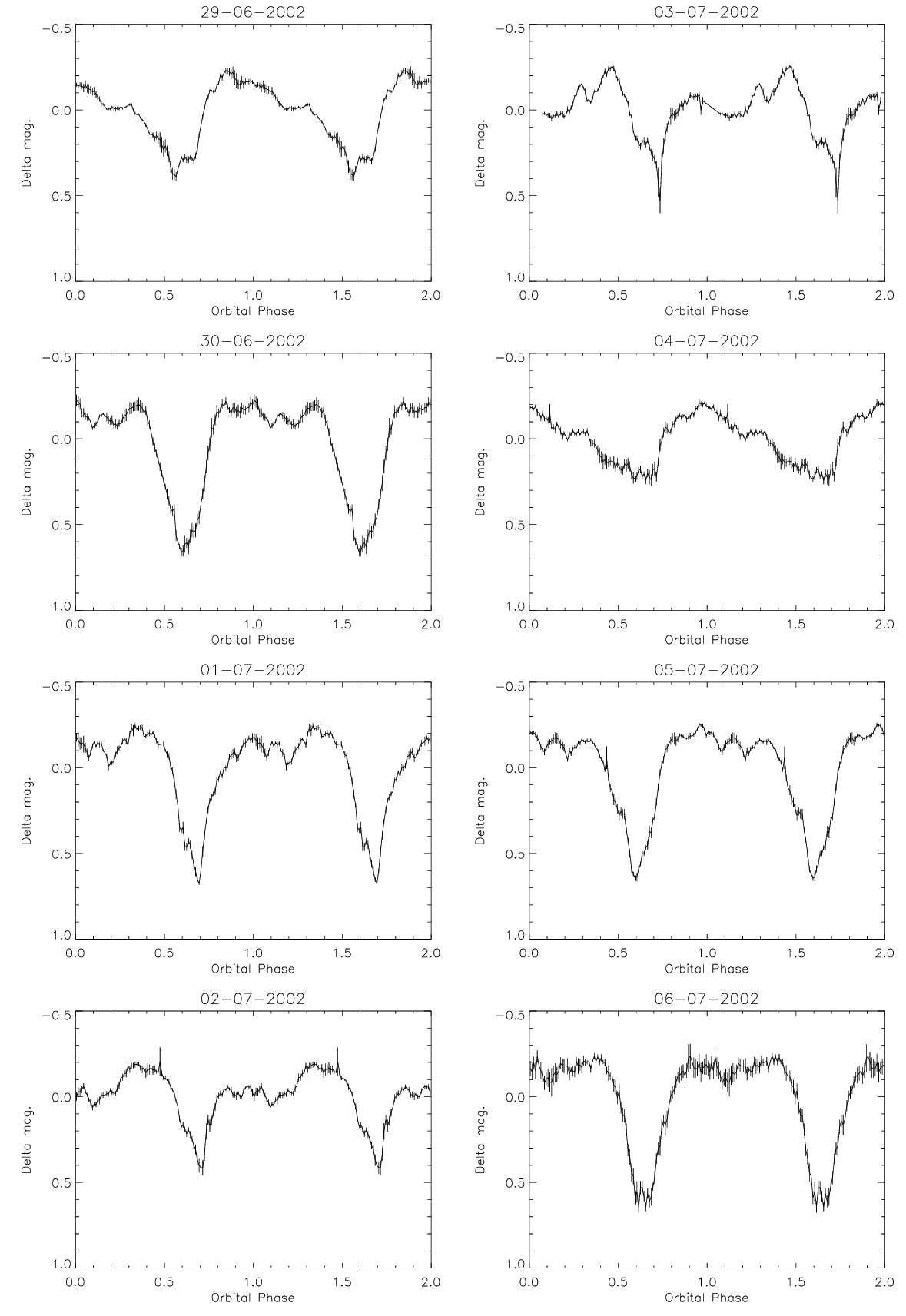}
 \caption{The June/July 2002 data set. Same as Figure 1.}
\end{figure*}

%\begin{table}
%  \centering
%\begin{tabular}{cc}
%\hline
%\hline
%Eclipse \# & HJD-2400000.0 \\
%\hline
% after \\ : \hline or \cline{col1-col2} \cline{col3-col4} ...
%1   & 47670.8031  \\
%2  &  47670.8817  \\
%3   &  47670.9582 \\ 
%13 &  47671.7297  \\
%14 &  47671.8078  \\
%26 &  47672.7313  \\
%27 &  47672.8101   \\
%29 &  47672.9637   \\
%39  & 47673.7336    \\
%40  &  47673.8114  \\
%\hline
%47130 & 51304.6025 \\
%47131  &  51304.6789  \\
%47143  &  51305.6054  \\
%47156  &  51306.6050  \\
%47157  &  51306.6826  \\
%47169  &  51307.6066  \\
%47170   &  51307.6841  \\
%47182  & 51308.6057   \\
%47183  & 51308.6831   \\
%47195  &  51309.6107  \\
%47196  &  51309.6883  \\
%\hline
%62056  &  52455.4385  \\
%62057  &  52455.5168  \\
%62069  &  52456.4435  \\
%62070  &  52456.5203  \\
%62083  &  52457.5248  \\
%62095  &  52458.4529  \\
%62096  &  52458.5269  \\
%62134  &  52461.4518  \\
%62135  &  52461.5274  \\
%62147  &  52462.4548  \\
%62148  &  52462.5333  \\
%\hline
%\hline
%\end{tabular}
%  \caption{ Eclipse timings of UW CrB, The first dataset id from Morris et al. (1990),
% the second is our 1999 run and the last our 2002 run.}\label{ }
%\end{table}

\begin{figure*}
% \centerline{\psfig{figure=lc1999.ps,height=20.0,width=18.0,angle=0}}
\includegraphics[scale=0.8,angle=90]{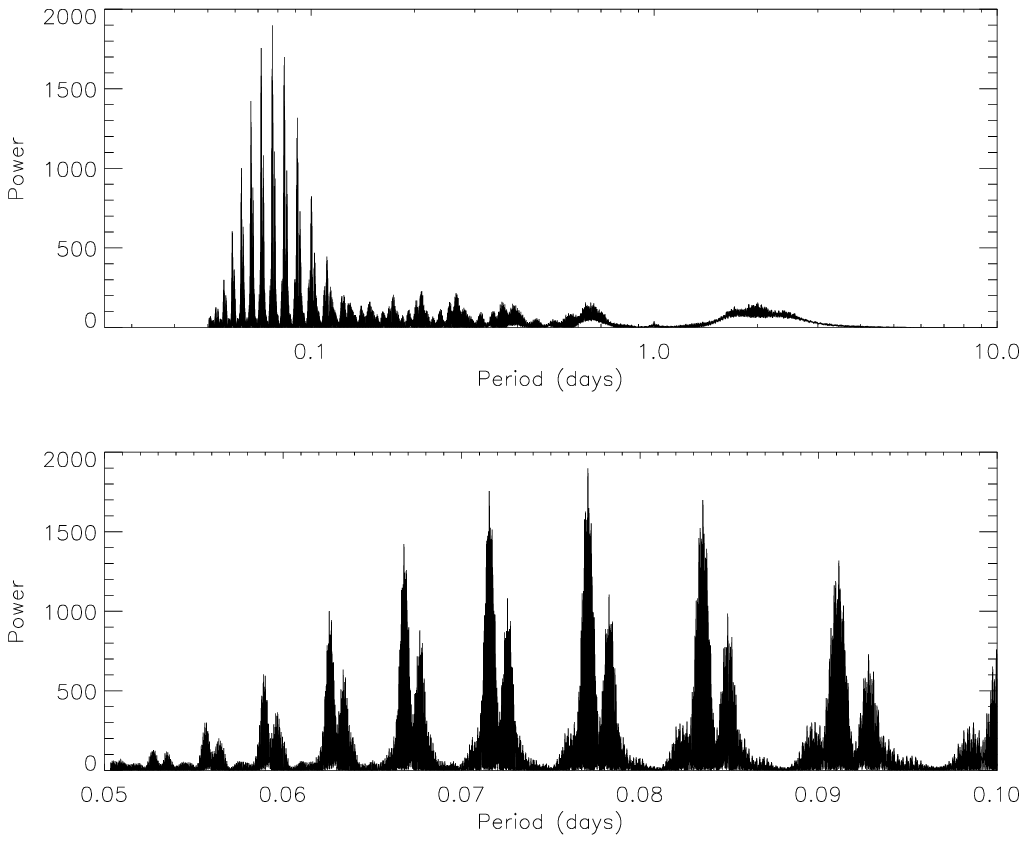}
 \caption{The power spectrum analysis of all the data (log-linear plot, top) and the same 
plot zoomed in around the best period of 0.077103.}
\end{figure*}

\begin{figure*}
% \centerline{\psfig{figure=lc1999.ps,height=20.0,width=18.0,angle=0}}
\includegraphics[scale=0.7]{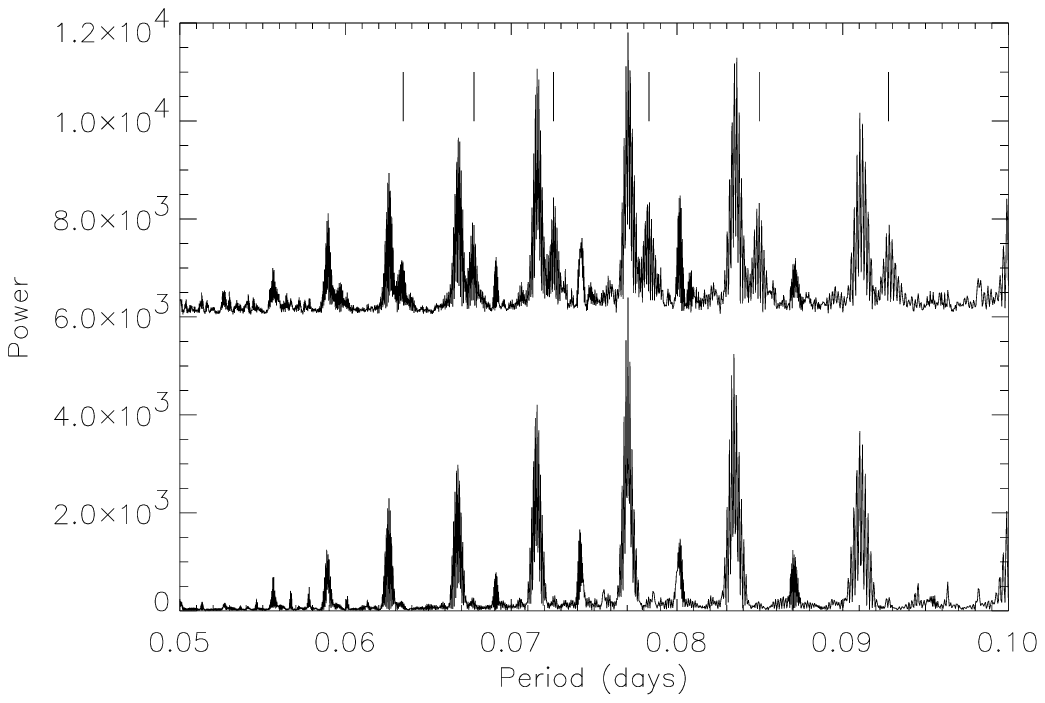}
 \caption{Epoch folding search periodogram around the best period. The top plot shows
 the periodogram of the data with the beat period aliases marked with vertical tick marks.
 The lower plot is a periodogram from simulated data having a similar average (but constant!) pulse shape as the real data. i.e. it is a representation of a simulated window
 function. Note that the peaks marked by the vertical lines in top plot are missing. See main text for discussion.}
\end{figure*}

\section{Data Analysis}

\subsection{Period Analysis}

Given the nature of the pulse profiles in UW CrB, it is far from clear that traditional astronomical period analysis methods
such as power spectral analysis or epoch folding search perform correctly. All these methods rely on the fact that 
the signal to be found is both stable in period and in pulse shape. As this is not entirely the case here, the resulting periodograms must be viewed with caution. 

We have chosen to tackle our data using various techniques in order to remove any method-dependent bias from
our results. This is also sensible, since the changing light curve pulse shapes present a challenge to any period
analysis method. We have started the ``usual" way,  i.e. ignoring the changes in the light curve shape, we have
computed the Lomb-Scargle power spectral density periodogram (Scargle 1982). This is shown in Figure 3.
The top panel shows the power vs. log(period) over a wide range of periods, and the lower panel zooms in near
the 111 minute period. The only periods present here seem to be the 0.07710315 d modulation, its second harmonic
(due to the secondary minimum in some light curves) and another spike at roughly 0.0783 d. Multiples of these spikes
are seen due to the very sparse sampling of the data in time (i.e the window function). 

Next, we proceeded with an epoch folding search, where the data are phase binned and then a null hypothesis of the source
being constant over the trial period is tested using a $\chi^2$ test (i.e. if the data do not vary over a given trial
period a low value of  $\chi^2$ is obtained, since the data are well represented by just a constant value over
the trial period).  This method is more sensitive to non-sinusoidal pulses, since increasing the number of phase
bins will allow for sharper features to be detected. In our analysis we have used 30 phase bins.

Figure 4 (top) shows the resulting periodogram (period vs. $\chi^2$) from our epoch folding search algorithm.
Now the method correctly picks up the right spike (0.07710315 d) as the highest one. Again, there is severe
aliasing in the periodogram due to the 1 day gaps in the data and (not properly seen in Figure 4) due to the shortness
of the period and the very long 3.5 year gap between the two data sets. 

In order to further investigate the effects of the sampling on our periodogram we have performed a simulation.
First, we have modelled the average light curve of UW CrB by fitting a 6th order Fourier series to the data. Having
done this, we were able to create a synthetic light curve with exactly the same sampling (time points) and similar
period and pulse shape to our real data. This was done by adding a similar amount of noise to the fit as there
was in the original light curve. We then proceed to perform exactly the same epoch folding search on the synthetic
data that we did on the real data. The resulting synthetic periodogram is plotted under the real one in Figure 4. This
plot is a true reproduction of the effects of the sampling (window function) and binning used in the epoch folding
search on the data that has a period and average pulse shape like those of UW CrB. We can see that most of the
features are reproduced, except for the spikes marked by short vertical lines in the upper panel of Figure 4. 
We investigated these spikes further, and they turned out to be offset from the spikes corresponding to
the 0.07710315 d period (and its window-related aliases) by a constant frequency of $\sim$ 0.2 cycles/d, i.e.
they correspond to a period of  $\sim$ 0.0783 d. We will return to this period later in this paper, but it is interesting
to note that this would be the synodic (beat) period if the system would have another longer period of 5 days.
This could be a disc precession period and would then imply prograde precession. In other words the 0.0783 d period would be a positive superhump period.

 \begin{table*}
  \centering
\begin{tabular}{ccccccccccccccc}
\hline
\hline 

%after \\ : \hline or \cline{col1-col2} \cline{col3-col4} ...

Nights  & 1     & 2      & 3      & 4      & 5      & 6      & 7      & 8      & 9      & 10     & 11     & 12     & 13     & 14      \\
\hline
  1         &  1.0  &    	&	  &	    &          &         &          &         &         &           &           &          &           &            \\
  2         &   0.939 & 1.0 &	  &	    &          &         &          &         &         &           &           &          &           &            \\
  3         &   0.817 &  0.825	& 1.0  &	    &          &         &          &         &         &           &           &          &           &            \\
  4         &   0.795 &  0.836 & 0.642 & 1.0  &          &         &          &         &         &           &           &          &           &            \\
  5         &   0.923 &  0.909 & 0.759 & 0.875 & 1.0   &         &          &         &         &           &           &          &           &            \\
  6         &   0.895 &  0.905 & 0.722 & 0.716 &  0.893 & 1.0  &          &         &         &           &           &          &           &            \\
\hline
  7         &   0.742 &  0.795 & 0.788 & 0.911 &  0.776 &  0.610 & 1.0   &         &         &           &           &          &           &            \\
  8         &   0.904 &  0.910 & 0.769 & 0.946 &  0.940 &  0.857 &  0.928 & 1.0  &         &           &           &          &           &            \\
  9         &   0.925 &  0.923 & 0.757 & 0.773 &  0.932 &  0.959 &  0.679 &  0.905 & 1.0  &           &           &          &           &            \\
  10         &   0.910 &  0.874 & 0.629 & 0.637 &  0.868 &  0.937 &  0.487 &  0.803 & 0.933 & 1.0    &           &          &           &            \\
  11        &   0.899 &  0.833 & 0.606 & 0.526 &  0.765 &  0.869 &  0.412 &  0.733 & 0.876 &  0.931 & 1.0    &          &           &            \\
  12         &   0.656 &  0.730 & 0.772 & 0.842 &  0.719 &  0.563 & 0.927  & 0.848 & 0.650 &  0.462 &  0.387 & 1.0   &           &            \\
  13         &   0.893 &  0.909 & 0.793 & 0.940 &  0.933 &  0.838 &  0.911 &  0.984 &  0.874 &  0.751 &  0.657 &  0.875 & 1.0    &            \\
  14         &   0.926 &  0.942 & 0.779 & 0.885 &  0.942 &  0.928 &  0.808 &  0.958 &  0.957 &  0.885 &   0.849 &  0.740 &   0.941 &  1.0    \\ 
\hline
\hline
\end{tabular}
  \caption{The correlation coefficients between all the possible pairs of nights. Night 1--6 correspond to the 'May 1999' , whilst  
		nights 7--14 are the 'July 2002' nights. }\label{ }
\end{table*}

\subsection{Light curve shapes}

We have observed UW CrB on 19 different nights altogether. During all of these nights we detect a clear modulation at the
111 minute period. However, there is striking night-to-night variability in the pulse shape of the light curve. This is not entirely
random though, since the light curve minimum seems to be relatively stable in phase. Another striking feature in the light
curves is that the light curve pulse shape seems to repeat roughly at a period of 5 days. We first noted this in the 1999 
data (Figure 1), where there is unfortunately only one pair of light curves that are 5 days apart. This prompted us to obtain
a second, longer observation (2002 data, Figure 2) that now contains 3 pairs of light curves separated by 5 days. 
Simply by visual inspection one can see that the first night's very broad asymmetric eclipse is roughly reproduced
during the sixth night, also the second and third nights' light curves resemble those of the seventh and eighth nights'. Next we proceed
to examine this apparent similarity in more detail.   

We have performed some statistical tests  in order to quantify the visually apparent 5 day periodicity in the 
modulation of the light curve shapes. To start with, we have correlated the folded light curves in pairs against
each other. The largest possible correlation coefficient was searched for using cross-correlation with about
0.05 maximum phase shift between the pairs of light curves (allowing for some error in period). As a result, quite high values
for the correlation coefficient were generally obtained (see Table 2 for the actual coefficients). This is a natural consequence of the fact that all light curves do
show a minimum, even if it is sometimes shallower and broader. Studying correlation alone, there is no evidence
for the 5 day cycle in the first data set (correlation between nights 1 and 6 is 0.895). However, the correlations between 
the pairs of nights 5 days apart in the second data set are amongst the highest in the whole table (0.927, 0.984 and 0.957). 
It is very difficult to asses the significance of these correlation values though. If we simply assume that our 91 correlation
values are a representative sample of correlation coefficients for any light curve pairs, we can sort the 
correlation coefficients and study the locations of the 4 pairs of nights that are 5 day apart from each other. From this it
follows that the correlation coefficients that we are interested in have ranks 1, 4, 17 and 32 among the 91 
coefficients. Now we can get a very conservative  \textit{upper limit} for the probability of this happening at random,
 i.e. the probability that 4 randomly selected coefficients (out of 91) would all be within the 32 highest. This upper limit is 
1.3 \% . In addition, we can define another test statistic by summing up the {\bf ranks} of 4 randomly chosen correlation
coefficients and comparing their distribution (approximately Gaussian, except for the wings) with the value (54) we get for our four pairs of nights separated by 5 days. As a result, we find that in only 0.47\% of cases the 4 randomly drawn pairs of nights have a summed correlation rank lower than 54. In conclusion, the distribution of correlation coefficient values, together with the values obtained for the 4 pairs of nights that are 5 days apart, support the idea that the light curve shape is modulated at approximately 5 d period.

Our second line of attack for measuring the significance of the 5 d cycle is based on cluster analysis of light curve
shapes. Here we have employed a method based on unsupervised neural networks called Self Organizing Maps (SOM)
Kohonen (1988) (See also Hakala et al. (2002) for a detailed description of the algorithm and recent usage of SOM in regularization). SOM is a topology-conserving clustering algorithm that is capable of finding (learning) intrinsic 
clustering in N-dimensional data. In our case we have used a 1-dimensional SOM network to search for 
similarities between light curves from different nights. 

For the clustering purposes we can think of a single 100 bin phase-folded light curve as a point in 100-dimensional space.
This means that (using July 2002 data only) our 8 folded and phase-binned light curves define 8 input data points 
for our SOM algorithm. Considering cluster analysis for 8 data points may seem unreasonable to start with, but
given some constraints we show that it does make sense.  

The idea is that if we define a maximum number of say 5 cluster centres, then, as we observe 5 light curve
shapes over the 5 d period, each cluster would represent the shape of the light curve at a different phase over 
that 5 d period. Now if we take our second data set that covers 8 nights, this means that in principle the first 5 nights' 
data could be mapped into different clusters, but the algorithm has to decide where to place the light curves of the 
remaining 3 nights. Now, IF there would be a 5 day cycle in the modulation light curve shape, the clustering algorithm should 
place the sixth light curve in the same cluster as the first, the seventh light curve in the same 
cluster as the second and the eighth light curve in the same  cluster that contains the third night's data.   
If there is no modulation, then the algorithm should spread the last three night's data evenly in the five clusters.
Thus, as SOM is an iterative stochastic algorithm and if we run this several times with different seeds for the random number generator, we can examine at what level the SOM algorithm finds evidence for the 5 d period (i.e. how often the data
separated by 5 nights gets classified into the same cluster).
Based on 100000 cluster analysis runs on our data we find that in 99.994\% of the
cases the algorithm comes out with a final clustering that places all three pairs of light curves observed five days apart in the same clusters. In order to further quantify this result
we performed some simulations on non-correlated data. These showed that the chance 
probability of such clustering is 21.6\%.

Finally, the ``synodic peaks" in Figure 4 that show a constant separation from the 
the main periodic peak (and its aliases) imply a synodic period of 0.0783 d, which in
turn, given the orbital period of 0.07710315 d, further implies a  superorbital period
of 5.04 d. Based on these three different pieces of evidence we conclude that the light curve shape is modulated at this period. The simplest explanation for such modulation
is either accretion disc precession or other periodic changes in the geometry of the disc.
We can identify the 
superhump period as 0.0783 d, whilst the 0.077103 d period is the orbital period. These,
together with the correlation analysis, suggest a precession period of $\sim$ 5 d for the
accretion disc. This enables us to estimate the system mass ratio using the empirical
formula for the superhump period excess (Patterson, 2001): q=0.216$\epsilon$, where  $\epsilon$=P$_{sh}$-P$_{orb}$/P$_{orb}$. In our case $\epsilon$=0.016 implies a mass ratio
q=0.072. Finally, as type I X-ray bursts have been detected from the source, we assume that the compact object is a neutron star. As such its most likely mass is 1.4M$_{\odot}$. This then implies M$_{2}\sim$0.1M$_{\odot}$ for the secondary, a factor
of two less than expected for a Roche lobe filling main sequence secondary at the given
period. We thus conclude that the secondary is likely to be somewhat evolved.

\section{Optical bursts}

As already noted earlier by Muhli et al. (2004) and Hynes et al. (2004), UW CrB shows
clear optical bursts that are most probably associated with the type I X-ray bursts seen
earlier (Mukai et al. 2001). They can be modelled by an instantaneous rise followed by an exponential decay (Hynes et al. 2004). These bursts are also well represented in our data set, where we can isolate 11 clear bursts. These are listed in Table 2. Typically the bursts have a peak amplitude equal to the out-of-burst flux and an e-folding time of  10--25 s. The total flux emitted during a burst corresponds roughly to the flux emitted by the rest of the system during 1--2 e-folding times (in optical). The mean errors for the amplitude (height), e-folding time and the fluence are (based on model fitting), 0.26, 0.1 s and 3.8 respectively. 

We observed a total of eleven bursts in our data set (Figure 5.). This means approximately one burst per 5.9h or 3.2 binary orbits. This is compatible with the estimate of one burst in
6.3h presented by Hynes et al. (2004) earlier (based on only five bursts). 

\begin{figure}
% \centerline{\psfig{figure=lc1999.ps,height=20.0,width=18.0,angle=0}}
\includegraphics[scale=0.7]{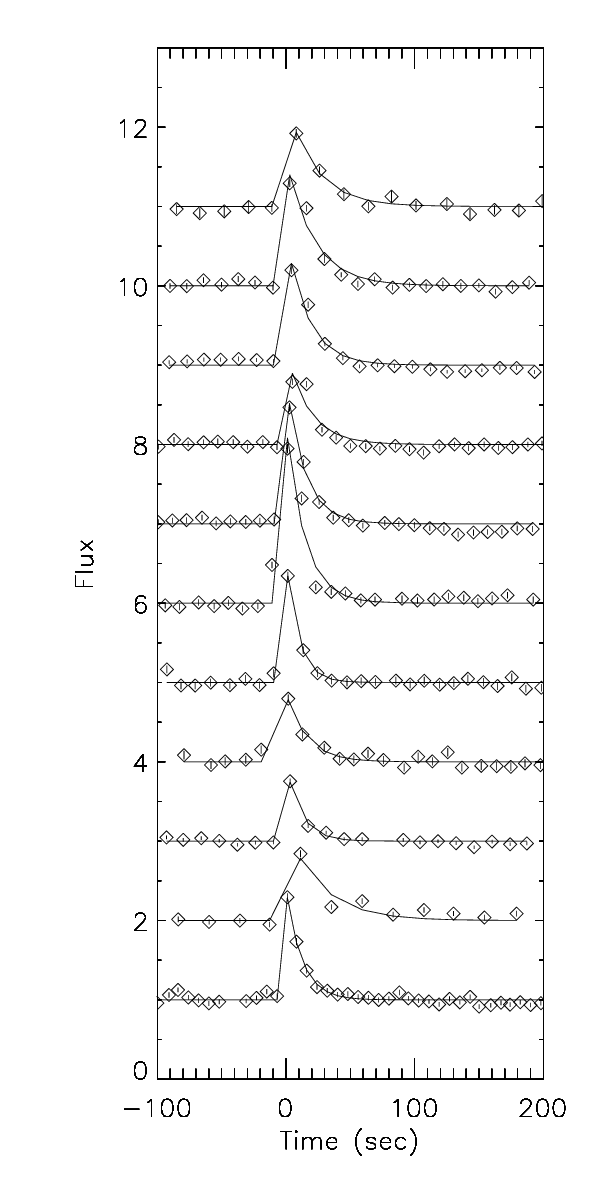}
 \caption{The individual burst light curves with model fits. The fit parameters are
 listed in Table 2.}
\end{figure}

The observed bursts have e-folding times in the range of 10--25 s and their
maximum amplitudes are not correlated with the e-folding times. We have also
looked at the orbital phases of the bursts. In order to increase our statistics we
have, in addition to our 11 bursts reported here, added another five bursts reported
by Hynes et al. (2004) and yet another five seen during the XMM-Newton observations
(Hakala et al. 2005), bringing the total number of bursts to 21. The burst distribution as a function
of orbital phase is plotted in Figure 6. Interestingly the bursts are concentrated on the
first half of the orbit and 16 out of the 21 detected bursts occurred in the orbital phase range 0.0--0.5. This is
in agreement with the view that the accretion disc is thicker (or more warped) on 
the side where the accretion stream impacts the disc, i.e. the disc prevents us from
seeing at least some of the bursts that occur at late orbital phases. Since our sample
is only 21 events, we have carried out simulations in order to investigate the significance
of the asymmetry in the burst phase distribution. As a result we conclude that if the burst phase
occurs at random, then we would see 16 or more bursts at phase range 0.0--0.5 only in 1.3\% of cases. 
However, if we allow the 16 or more bursts happen in any phase
range of length 0.5, then the chance probability increases to 20.8\% .More burst 
detections are clearly needed, but based on the distribution we have, it is possible that
the disc is asymmetric and that the true burst frequency is higher than reported here.  

It is interesting to note that all of the bursts modelled in this paper (Fig. 5. and Table 3.) seem to
be rather similar in flux irrespective of their orbital phase. This is curious, since if we believe that the optical
bursts are a result of reprocessing of X-ray bursts in a warped/vertically extended asymmetric disc and/or
the secondary, we would expect that the bursts seen during the 0.5--1.0 phase range should be fainter, as
there is i) more obscuring matter, i.e. the disc bulge, obscuring the view of the inner disc and ii) less projected 
area in the plane of sky available for reprocessing (assuming there is less vertical extension on the other edge
of the disc). However, there is no clear evidence for any fainter bursts in our data. One possible explanation
could be that maybe the optical bursts are produced by the enhanced optical emission due to the 
reprocessing in the inner disc. In this case it would be obvious that sometimes we simply don't see the reprocessing
area at all and that this could happen more often in the 0.5--1.0 phase range, where the disc is generally more elevated.
Yet the changing disc structure would occasionally enable us to see some bursts at any orbital phase. The key test would be simultaneous high time resolution X-ray and optical observations of the bursts in order to measure the possible reprocessing lags. However, since UW CrB is most likely an ADC source (and faint both in X-rays and optical), where we believe that the NS is not seen directly, observing such lags might be almost impossible.       

\begin{figure}
% \centerline{\psfig{figure=lc1999.ps,height=20.0,width=18.0,angle=0}}
\includegraphics[scale=0.5]{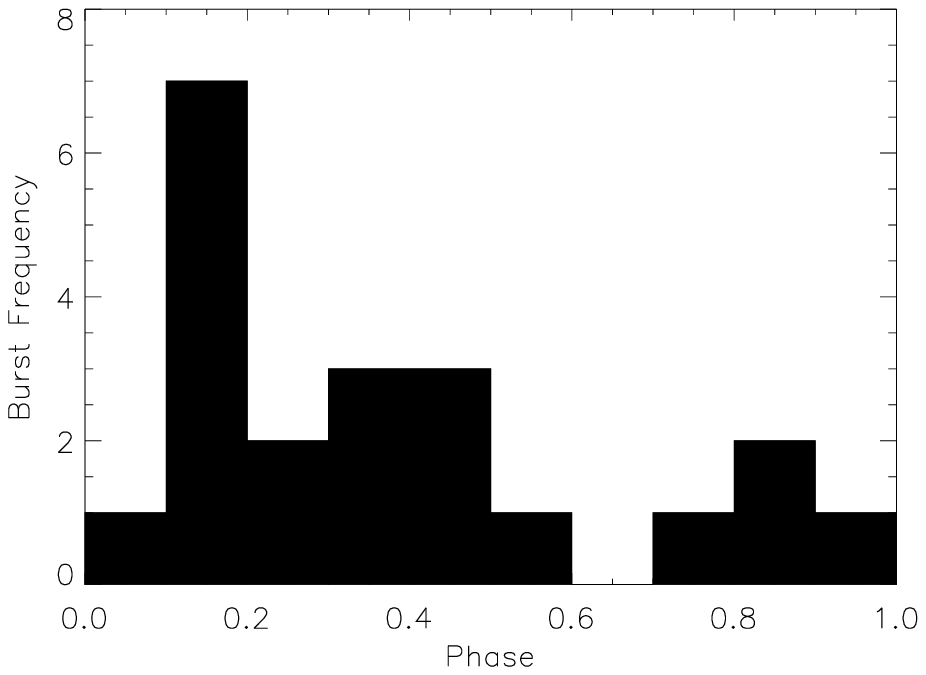}
 \caption{The orbital phase distribution of the 21 bursts reported in total.}
\end{figure}

\begin{table}
\caption{The observed optical bursts. The amplitude and flux units
are arbitrary.}
\begin{center}
\begin{tabular}{|c|c|c|c|}
\hline
HJD & Amplitude & e-folding time (sec) & Flux \\   
\hline
       2451213.747   &   1.38   &   22.3    &  27.2 \\
       2451214.689   &  1.61    &  21.3     & 30.5 \\
       2451215.767   &   1.67   &   16.9    &  25.0 \\
       2451304.614   &   1.21   &   17.9    &  19.1 \\
       2451304.698   &   1.87   &   14.2    &  23.5 \\
       2451307.613   &   2.26   &   14.5    &  29.0 \\
       2451308.568   &   1.56   &   9.8    &  13.6 \\
       2451309.637   &  0.87  &    16.0   &   12.4 \\
       2452461.438   &   1.03   &  11.7     & 10.7 \\
       2452462.475   &   1.22   &   27.3    &  29.5 \\
       2452492.422   &   1.41   &   12.2    &  15.3 \\
\hline
\end{tabular}
\end{center}
\label{default}
\end{table}

\section{Conclusions}

It is rather obvious from our photometry, that UW CrB is most likely a high inclination system, where at least a partial eclipse of an accretion disc is always visible, and that the bulk of the optical orbital modulation is caused by either varying self-shadowing and projected area of a non-axisymmetric accretion disc and/or a partial eclipse of a such disc by the secondary star. Hakala et al. (2005) noted that the X-ray light curves can be
best modelled by varying partial covering of the X-ray emitting region over the orbital
period. This suggests that the accretion disc either has substantial asymmetric vertical
structure or it is clearly warped out of the orbital plane. Our current data cannot
distinguish between these models. Our interpretation of the optical light curves does not make
any assumptions on the nature of the compact object itself. It is still worth noting that precession of accretion discs is a 
well known phenomenon in high mass ratio systems (Whitehurst, 1988). Given the magnitude of effects seen in UW CrB, it
is possible that this system is an extreme example of the superhump phenomenon seen in nonmagnetic cataclysmic variables
with similar orbital periods, i.e. SU UMa systems (Patterson, 2001).

\section*{Acknowledgments}

Just days before submission, an article by Mason et al. (2008) appeared on-line reporting results similar to ours. Our analysis independently confirms their findings on the 5d cycle related to the morphology changes in the accretion disc. We would like to thank Prof. Phil Charles and Dr. Osmi Vilhu 
for very useful comments and discussions. We would also like to thank
the Nordic Optical Telescope staff for enabling the flexible scheduling required
to cover the time baselines required. The data presented here have been taken using ALFOSC, which is owned by the Instituto de Astrofisica de Andalucia (IAA) and operated at the Nordic Optical Telescope under agreement between IAA and the NBIfAFG of the Astronomical Observatory of Copenhagen. The authors gratefully acknowledge support from the
Academy of Finland. PH and DCH have been supported by the Academy of Finland research fellowships. Finally, we would like to thank the anonymous referee for very constructive criticism that helped to improve the paper considerably.

\end{document}